\documentclass{pasj00}
\draft

\usepackage{natbib}

\newcommand{\ve}[1]{\mbox{\boldmath${#1}$}}
\newcommand\bfn{\mbox{\boldmath $\nabla$}}

\begin{document}
\SetRunningHead{Author(s) in page-head}{Running Head}

\title{Generation of Seed Magnetic Fields in Primordial Supernova Remnants}

\author{Hidekazu \textsc{Hanayama}\altaffilmark{1} %
}
\email{hanayama.hidekazu@nao.ac.jp}

\author{Keitaro \textsc{Takahashi}\altaffilmark{2}}
\email{keitaro@a.phys.nagoya-u.ac.jp}
\and
\author{Kohji \textsc{Tomisaka}\altaffilmark{1,3}}
\email{tomisaka@th.nao.ac.jp}

\altaffiltext{1}{National Astronomical Observatory of Japan, 
2-21-1 Osawa, Mitaka, Tokyo, 181-8588, Japan} 
\altaffiltext{2}{Department of Physics and Astrophysics, 
Nagoya University ,Chikusa-ku Nagoya 464-8602 Japan}
\altaffiltext{3}{Department of Astronomical Science,
The Graduate University for Advanced Studies [SOKENDAI], 
2-21-1 Osawa, Mitaka, Tokyo, 181-8588, Japan}

%


\maketitle

\begin{abstract}
Origin of the magnetic field ubiquitous in the Universe is studied
 based on the Biermann mechanism, which is expected to work in 
 the non-barotropic region. 
We perform a series of two-dimensional MHD simulations of the
 first generation supernova remnant (SNR) expanding in the
 inhomogeneous interstellar matter (ISM) and study the Biermann 
 mechanism working in the interior of the SNR. 
Especially, we pay attention to the relaxation process of electron
 and ion temperatures via the Coulomb interaction. 
In the early SNR in which the electron temperature is much lower
 than the ion temperature, the Biermann mechanism is ineffective,
 since the gradient of electron pressure is small. 
Magnetic fields begin to be generated just behind the shock front
 when the electron temperature is sufficiently relaxed.  
Assuming the explosion energy of $10^{52}~{\rm erg}$, the total
 magnetic energy generated reaches about $10^{26}~{\rm erg}$ and
 does not depend strongly on the parameters of either SNR or ISM. 
Analytic expression to estimate the magnetic total energy is
 presented and it is shown this agrees well with our numerical
 results. 
Finally we evaluate the expected amplitude of magnetic fields in
 protogalaxies as $\sim 10^{-19}~{\rm G}$, which is sufficient
 for seed fields of the observed galactic magnetic fields.
\end{abstract}

\section{Introduction}

The origin of magnetic fields in the universe has been an active
 field of modern cosmology and astrophysics \citep{wid02,kul08}.
Concerning the origin of galactic magnetic fields, which are
 observationally known to be of order $1 \mu {\rm G}$, it is
 argued that dynamo mechanism could amplify and maintain magnetic
 fields if there were some tiny fields ($\sim 10^{-20}~{\rm G}$)
 at the galaxy formation \citep[for example,][]{les95}.
There have been many approaches to the origin of the seed fields
 such as cosmological scenarios \citep{tur88,bam04,tak05,ich06}
 and generation via reionization \citep{gne00,lan05}. 
These scenarios predict that the universe is filled with very tiny
 magnetic fields about $10^{-15} - 10^{-25}~{\rm G}$ and
 there is a possibility that they could be observationally
 probed through the detection of delayed high-energy emission from
 gamma-ray bursts \citep{pla95,ich08,tak08}.

In this article, we focus on magnetogenesis in primordial
 supernova remnant (SNR) via the Biermann mechanism \citep{bie50}.
The Biermann mechanism was also studied in other systems such
 as the formation of protogalaxies \citep{kul97} and Pop III stars
 \citep{xu08}.
The Biermann mechanism is an induction process of magnetic
 fields and can be derived by combining a generalised
 Ohm's law with the Maxwell equation \citep{wid02},
\begin{equation}
\frac{\partial{\ve B}}{\partial t}
- \bfn \times \left( {\ve v}_p \times {\ve B} \right)
= \frac{m_e c}{e}
  \frac{\bfn P_e \times \bfn \rho_e}{\rho_e^2},
\label{indeqe}
\end{equation}
 neglecting the electric resistivity. 
Here, ${\ve v}_p$, $m_e$, $e$, $c$, $P_e$, and $\rho_e$ are the
 velocity of protons, the mass of electrons, the elementary
 electric charge, the light velocity, the pressure of
 electrons, and the mass density of electrons, respectively. 
When we assume $n_e \simeq n_p \simeq n$, $P_e \simeq P/2$, and
 ${\ve v}_p \simeq {\ve v}$ where $n_e$, $n_p$, $n$, $P$, and
 ${\ve v}$ represent the number density of electrons, protons, and
 ions, the total gas pressure, and the fluid velocity, respectively,
 equation (\ref{indeqe}) can be rewritten as
\begin{equation}
\frac{\partial{\ve B}}{\partial t}
= \bfn \times \left( {\ve v} \times {\ve B} \right)
  + \alpha \frac{\bfn P \times \bfn \rho}{\rho^2}.
\label{indeq}
\end{equation}
The coefficient $\alpha$ is defined as $\alpha \equiv m_p c/2e
 \simeq 0.5 \times 10^{-4}~{\rm G ~ sec}$, where $m_p$ is the mass
 of protons.

The Biermann mechanism requires a vorticity of plasma gas.
The vorticity is generated in the region where the spatial
 gradients of the pressure and density are not parallel.
There are some studies of the interaction of the shock of
 SNR and interstellar clouds \citep{kle94,naka06}. 
These studies revealed that the interaction produces the
 vorticity efficiently in the shocked region. 
The interaction of inhomogeneous interstellar medium and the
 shock of SNR also drives the turbulent motion of the gas,
 and the vorticity is generated in the bubble of SNR \citep{bal01}. 
Therefore, primordial SNRs occurred before the galaxy formation
 can be a candidate of the origin of galactic magnetic fields. 
Furthermore, it is important to note that primordial supernova
 explosions play an important role in the magnetogenesis,
 since the initial mass function (IMF) of population III stars
 is believed to be substantially top-heavy \citep[e.g.,][]{abe02}. 

For the generation of the seed fields by the primordial
 supernova explosions, \citet{mir98} studied the process
 with multiple explosion scenario of Population III objects. 
Although their situation considered is more or less similar
 to ours, their calculations are based on several ad hoc
 assumptions. 
First, they calculate magnetic field strength assuming a constant
 generation rate which is actually dependent on the configuration
 of the density and pressure gradients.
Second, they considered multiple SN explosions, and most of
 the magnetic fields is produced by the hypothetical objects
 with a mass $M_0=10^6-10^{10}~M_\odot$, and explosion energy
 $E_0=10^{56}-10^{60}~{\rm erg}$. 
This assumption seems inconsistent with the current understanding
 of star formation.

In our previous article \citep{han05}, we studied the generation
 of magnetic fields by the Biermann mechanism with realistic
 two-dimensional magnetohydrodynamic (MHD) numerical simulations. 
A single SNR with the explosion energy $10^{53}~{\rm erg}$ was put
 in inhomogeneous ISM and the generation and evolution of magnetic
 fields were followed.
Then we found the total energy of the magnetic fields to be
 $10^{28} \sim 10^{31}~{\rm erg}$ depending on the supernova and
 ISM parameters.

In this article, we present more detailed analysis especially
 focusing on the temperature relaxation between ions and electrons.
When we derived equation (\ref{indeq}) from equation (\ref{indeqe}), we used
 $P_e \simeq P$ which comes from the assumption that the ions and
 electrons have the same temperature, $T_e \simeq T_p$. 
In fact, the temperatures of ions $T_i$ and electrons $T_e$ just
 behind the shock differ each other and the Rankine-Hugoniot
 shock jump condition indicates
\begin{eqnarray}
&& T_i = \frac{3}{16} \frac{m_i v_{\rm bub}^2}{k_{\rm B}},
\label{jumpi} \\
&& T_e = \frac{3}{16} \frac{m_e v_{\rm bub}^2}{k_{\rm B}},
\label{jumpe}
\end{eqnarray}
 where $v_{\rm bub}$ is the shock velocity. 
Noting that the time scale of the Coulomb interactions between
 electrons and ions \citep{spi62} is longer than the age of SNR
 in the early phase, there is a possibility that the electron
 temperature behind the SNR shock front is much lower than the
 ion temperature.
In this case, the analysis with equation (\ref{indeq}) would
 overestimate the generation of magnetic fields.

For a long time, the temperature relaxation of ions and electrons
 behind the shock of SNR has been actively discussed by many
 authors \citep{ito78,cox82,cui92,masai94}. 
According to \cite{cox82}, in the unequilibrated region in the
 early phase of SNR the electron temperature is spatially
 isothermal. 
If this is the case, the gradient of electron pressure is
 negligible and, therefore, the Biermann mechanism is there
 ineffective. 
Thus, we have to follow the electron temperature in order to
 analyze the generation of magnetic fields correctly. 
This is the main ingredient of the current article.

On the other hand, in the radiative cooling phase, although
 the electron temperature is expected to be sufficiently
 equilibrated to the ion temperature, the radiative cooling
 process increases its importance with the decrease of
 the temperature at the shock front, and the thermal structure
 of the SNR becomes isothermal again \citep{sla92}. 
Then, the Biermann mechanism does not work effectively in the SNR.
Therefore, in this study, we focus on the generation of
 the magnetic fields  before the radiative cooling phase. 

In section \ref{sec:model}, the basic equations for numerical
 simulation of primordial SNRs, the numerical method, and
 the initial condition of SNR and its surrounding environment
 are given. 
In section \ref{sec:result}, we present the results of numerical
 simulations. 
We discuss the analytic estimations of the generated magnetic
 fields in section \ref{sec:discuss}.

\section{Models and Methods}\label{sec:model}

\subsection{The Basic Equations}
\label{sec:be3}

The basic equations for our numerical simulations are as follows:
\begin{equation}
\frac{\partial \rho}{\partial t} + \bfn \cdot (\rho \ve{v}) = 0,
\label{mass2}
\end{equation}
\begin{eqnarray}
\frac{\partial\rho \ve{v}}{\partial t}
   + \bfn \cdot (\rho \ve{v} \ve{v}) 
= - \bfn P
     - \bfn \left( \frac{{\ve{B}}^2}{8 \pi} \right)
     + \frac{1}{4 \pi} (\ve{B} \cdot \bfn) \ve{B},
\label{mome2}
\end{eqnarray}
\begin{eqnarray}
 \frac{\partial}{\partial t}
   \left(E + \frac{{\ve{B}}^2}{8 \pi}\right) 
 + \bfn \cdot
     \left[ (E + P) \ve{v}+ \frac{1}{4 \pi} 
     \left\{ \ve{B} \times (\ve{v} \times \ve{B}) \right\}
     \right] = 0,
\label{ener2}
\end{eqnarray}
\begin{equation}
\frac{\partial \ve{B}}{\partial t}
= \bfn \times (\ve{v} \times \ve{B})
  + \alpha \frac{\bfn P \times \bfn \rho}{\rho^{2}},
\label{ind2}
\end{equation}
 where $\rho$, $\ve{v}$, $E$, $P$, $\ve{B}$, and $T$ are the density,
 velocity, total energy, pressure, magnetic field, and temperature, 
 respectively. Here the total energy of the gas is defined as
\begin{equation} 
E = \frac{1}{2}\rho |\ve{v}|^2 + \frac{P}{\gamma - 1},
\end{equation}
 where we fix the value of the adiabatic index $\gamma$ to $5/3$,
 which is valid in case of the non-relativistic gas. 
The last term of the right-hand side of equation (\ref{ind2}) 
 is called Biermann term and expresses the Biermann effect,
 where $\alpha = m_{p}c/2e$. We assumed the gas is fully ionized
 and the number ratio of H to He to be $10:1$. 
Then the mean molecular weights are $\mu_0=14/11=1.27$ for nuclei
 and $\mu_t=14/23=0.61$ for fully-ionized ions and electrons.
The number density of the hydrogen atoms is $n_{\rm H}=(10/11)n$
 where $n$ means the number density of the ions
 ($n_{\rm H}+n_{\rm He}$). 
Since we focus on the generation of the magnetic fields in
 adiabatic expansion phase, we ignore the radiative cooling.

\subsubsection{Generation Region of Magnetic Fields}
\label{sec:gf2}

We pay attention to the region where the magnetic fields is
 generated by the Biermann mechanism. 
The interaction between the shock wave and the inhomogeneous ISM
 generates the vorticity. 
Therefore, especially, in the transition layer of the shock front,
 the baroclinic term $(\bfn P \times \bfn \rho)/\rho^2$ is expected
 to work effectively because the gradient of the pressure is much
 larger than in other regions. 
However, the thickness of the layer is much smaller compared with
 the radius of SNR, and is on the order of, or larger than the ion
 inertia length of $\sim 10^7/n^{1/2}~{\rm cm}$ \citep{ito84}.
In such a region, although the behaviour of the plasma particle
 must be considered, it depends on complex processes of
 the interaction of ions and electrons, and is conventionally
 studied by particle simulations \citep{shima00,kato08}.
If the electron pressure $P_e$ is not as high as the proton 
 pressure $P_p$ there, the Biermann mechanism may not work
 effectively. 
In fact, \cite{shima00} predicted the ratio between the electron
 and the ion temperatures $T_e/T_i \sim 0.2$ in the transition
 layer with Alfv$\acute{{\rm e}}$nic Mach number of
 $M_{\rm A} = 20$. Observationally, the ratio $T_e/T_i$ is less
 than $\sim 0.1$ in the transition layer with a shock velocity
 larger than $\sim 2000~{\rm km~s^{-1}}$ \citep{rako05}. 
Since in the transition layer the electron temperature is
 expected to be lower than the ion temperature, we ignore the
 generation process working in the transition layer as long as
 we consider the early phase of SNR. 
That is, we only take the Biermann effect in the interior of SNR
 into account, even if it would lead to an underestimate of
 magnetic fields.
Thus, our analysis is rather conservative.

In our calculation we introduce a switch $C_{\rm sw}$ that
 discriminates the region of the post shock region where we
 calculate the Biermann term $(\alpha \bfn P \times \bfn \rho)/\rho^2$ 
 in our one-temperature/one-pressure description. 
For the search of the boundary between the transition layer and the
 post shock region, the switch $C_{\rm sw}$ is defined as follows
 : $C_{\rm sw}=-1$ in the region of $(\bfn P) \cdot \ve v < 0$
 ; $C_{\rm sw}= 1$ in the region of $(\bfn P) \cdot \ve v > 0$
 ; $C_{\rm sw}= 0$ in the region of $(\bfn P) \cdot \ve v = 0$.
If the switch $C_{\rm sw}$ is equal to $0$ or $-1$, then we reduce
 the Biermann term to zero. 
In this way, we discuss the generation process of the magnetic
 fields in the inner region of SNR.
Furthermore, since a region with $T_e/T < 0.99$ has isothermal
 $T_e$ distribution, magnetic generation is ignorable there.
Therefore, we restrict the generation region of the magnetic 
 fields to the postshock region with $T_e/T \equiv g \ge 0.99$. 
Thus, our calculation must give a same result obtained with
 two-temperature simulation effectively.
Then, the induction equation (\ref{ind2}) with the relaxation
 effect of the electron temperature is written as
\begin{eqnarray}
\left\{
\begin{array}{l}
{\displaystyle \frac{\partial \ve{B}}{\partial t}}
= \bfn \times (\ve{v} \times \ve{B})
  + \alpha {\displaystyle 
  \frac{\bfn (gP) \times \bfn \rho}{\rho^{2}}}, 
~~~~(g \ge 0.99\ {\rm and}\ C_{\rm sw} = 1) \\
{\displaystyle \frac{\partial \ve{B}}{\partial t}}
= \bfn \times (\ve{v} \times \ve{B}). 
~~~~~~~~~~~~~~~~~~~~~~~~~~~~({\rm others}) \\
\end{array} \right.
\label{ind3}
\end{eqnarray}

\subsubsection{Equilibrium Equation of Electron Temperature}
\label{sec:te2}

We also consider the relaxation of the electron temperature.
In the adiabatic expansion phase of SNR, the relaxation equation of
 electron temperature is derived from the energy equation for
 the electron gas with the heating effect driven by the Coulomb
 collision with ions. 
Here, we ignore the thermal conduction and the radiative cooling. 
The relaxation equation was given by \cite{ito78}, and we convert
 the formulation according to \cite{cox82},
\begin{equation}
\frac{df}{dt} = \frac{\partial f}{\partial t} + ({\ve v} \cdot \bfn) f
= \frac{\ln \Lambda}{80}
\left( \frac{n}{T^{3/2}} \right),
\label{relax32}
\end{equation}
 where $\ln \Lambda$ is the Coulomb logarithm and $f$ is a function
 of $g = T_e/T$ defined as
\begin{equation}
f = \frac{3}{2} \ln \left( \frac{1+g^{1/2}}{1-g^{1/2}} \right)
- g^{1/2} (g+3).
\label{relax33}
\end{equation}
After solving equation (\ref{relax32}) for $f$, $g = T_e/T$ can be
 obtained from an approximate formula
\begin{equation}
g \approx 1 - \exp \left[ -\left( \frac{5}{3}f \right)^{0.4}
\left\{ 1 + 0.3 \left( \frac{5}{3}f \right)^{0.6} \right\} \right].
\label{relax34}
\end{equation}
We solve equation (\ref{relax32}) with other MHD equations and obtain
 the electron temperature $T_e$ with using equation (\ref{relax34})
 and the gas temperature distribution $T$. 
Since the temperature ratio $g=T_e/T$ takes $1/1836$ just behind the
 transition region (see equations (\ref{jumpi}) and (\ref{jumpe})),
 we take a boundary condition as $g = 1/1836$ at the transition
 region ($C_{\rm sw} = -1$).

\subsection{Numerical Methods}
\label{sec:nm2}

We solve the above equations by a two-dimensional MHD code
 in the cylindrical coordinates $(r, z, \phi)$ assuming axial
 symmetry around the symmetry axis ($z$). 
Since we start the calculation from $\ve{B} = 0$, it is sufficient
 to consider only the $\phi$-component of the magnetic fields
 $B_{\phi}$ which is generated from the Biermann term.

The code is based on the MHD Roe scheme \citep{roe81,car97}
 coupled with MUSCL technique \citep{hir90} to achieve
 the second order spatial and temporal accuracy. 
We employ improved rotated-hybrid Riemann solvers \citep{nishi08}
 to overcome so-called Carbuncle phenomenon, which is a numerical
 instability appeared in the shock wave. 
Curing $\bfn \cdot \ve B$ error, we adopted the hyperbolic
 divergence cleaning given by \cite{ded02} and \cite{matsu07}. 
We note that equation (\ref{relax32}) is also solved by using the
 fully upwind scheme with MUSCL and Carbuncle cure technique which
 achieve the second order spatial-temporal accuracy and numerical
 stability.

The numerical scheme was tested by several problems. 
We checked the code by comparing with MHD shock tube problems
 \citep{bri88} and the adiabatic SNR with the analytic Sedov
 solution \citep{sed59}.
In high resolution calculation ($4000 \times 4000$ grids) of
 primordial SNR expanding into a uniform ISM (explosion energy $E_0$
 and ISM density $n_0$ are $10^{51}~{\rm erg}$ and
 $0.2~{\rm cm^{-3}}$, respectively), the postshock density peak value
 of the shock front at $1500~{\rm yr}$ after the SN explosion
 was $95.8\%$ of the analytical value (four times larger
 than the density of ISM), for example. 
In addition, we compared our results with those of \citet{naka06}
 for the case of interaction of a plane-parallel shock with
 a cloud to check the robustness of our calculation of the
 vorticity derived from the induction equation (\ref{ind2}). 
As for the normalized peak value of total
 circulation $\Gamma$ in the evolution of the cloud distracted
 by the planar shock, it is analytically estimated as $-1.77$
 \citep{kle94} for the model of AS8 \citep[see][]{naka06}.
The result of \citet{naka06} is $-2.01$ while that of our
 calculation is $-1.61$, and both results are in good agreement
 with Klein's estimation within 10\%.

In computation, our numerical domain covers a region of
 $128~{\rm pc} \times 128~{\rm pc}$ with $2048 \times 2048$
 grid points. 
The grid spacings are $\Delta r = \Delta z = 0.0625~{\rm pc}$
 and the length scale of density fluctuation ($8~{\rm pc}$,
 see below) is resolved with 128 grids. 
This is sufficient for the convergence of the total magnetic
 energy as we show later.
In the simulated region of $128~{\rm pc} \times 128~{\rm pc}$,
 the average density of ISM is approximately constant even
 in the dark halo structure \citep{kita05}.

\subsection{Initial Conditions}
\label{sec:ic3}

We assume inhomogeneous ISM with average number density of
 $n_{0} = 0.2~{\rm cm}^{-3}$ for initial conditions of our fiducial
 model, and also consider a model with $n_{0} = 0.8~{\rm cm}^{-3}$
 as a variation. 
These assumptions are consistent with the previous
 studies \citep{kita05,gre07}, in which they obtained initial
 conditions for Population III SN sites that were calculated by
 self-consistent radiative transfer calculations. 
The scale length of the density fluctuation and the amplitude of
 the inhomogeneity are not well understood. 
Recently, \cite{wis08} studied the formation process of HII region
 around a very massive first star with 3D numerical simulations,
 and showed the anisotropic expansion and inhomogeneous density
 structure in the HII region with the scale of $\sim 1~{\rm kpc}$. 
In their study, the HII region has relatively large density
 inhomogeneities and the spatial scale of the large inhomogeneity
 seems to be of the order $10 \sim 100~{\rm pc}$. 
Therefore, we assume here inhomogeneity with relative density
 fluctuation $\delta n \equiv (n-n_0)/n_0$ derived from a Gaussian
 spectrum with concentration around a scale length $\lambda_0$
 ($32$ and $64~{\rm pc}$) and amplitude $A_0$ ($0.25$ and $0.125$),
 where we assume the power spectrum with a peak at the wave number
 $k_0~(=128~{\rm pc}/\lambda_0)$ as 
 $A_0 \exp[-(k-k_{0})^2/2{\sigma_{k}}^2]/2\pi k$ and set the
 variance $\sigma_{k} = 0.5$ to achieve $-1 < \delta n < 1$. 
Avoiding the biased distribution, a normally-distributed
 random number in the range from 0.9 to 1.1 is multiplied to
 the amplitude of $\delta n$ in Fourier space and a
 uniformly-distributed random number in the range from 0 to
 $2\pi$ is added to the phase.
Then, the number density is given by using $\delta n$ in real
 space as $n = n_{0}(1+\delta n)$.

For the explosion energy of the supernova, we adopt
 $E_{0} =10^{52}~{\rm erg}$ for the fiducial model
 \citep{kita05,gre07}. 
This explosion energy corresponds to stars with mass $200~M_{\odot}$
 which explodes as a pair-instability supernova \citep{fry01}. 
Also we consider a model with $E_{0} = 2.5 \times 10^{51}~{\rm erg}$
 as a variation.

We begin the simulation by adding thermal energy of $E_0 = 10^{52}$
 or $2.5 \times 10^{51}~{\rm erg}$ to the cells near the origin
 $(r,z) = (0,0)$ with Gaussian distribution. 
Initial distribution of thermal pressure $P_{\rm SN}(r,z)$ with
 explosion energy $E_0$ is defined as
\begin{equation}
P_{\rm SN}(r,z)
= \frac{(\gamma-1) E_{0}}{(\sqrt{2\pi}\sigma_{\rm SN})^3}
  \exp{\left(-\frac{z^2+r^2}{2{\sigma_{\rm SN}}^2}\right)}, 
\label{eq:snini}
\end{equation}
 where ${\sigma_{\rm SN}}^2$ represents variance of Gaussian
 distribution and we assumed $\sigma_{\rm SN} = 2~{\rm pc}$.

We set the gas density uniform from the origin $(0,0)$ to
 $10~{\rm pc}~(= 5\sigma_{\rm SN})$ in the radius, assuming
 the effect of the intense radiation field and stellar wind from
 a progenitor star. 
In addition, to connect the homogeneous region and inhomogeneous
 region smoothly, $\delta n$ was replaced by $\delta n'$ where
\begin{equation}
\delta n'(x,y,z)
= \left(\frac{\sqrt{r^2+z^2}-5\sigma_{\rm SN}}
             {10\sigma_{\rm SN}-5\sigma_{\rm SN}}
  \right)^2
  \delta n(r,z),
\end{equation}
 in the region from $10~{\rm pc}~(=5\sigma_{\rm SN})$ to 
 $20~{\rm pc}~(=10\sigma_{\rm SN})$ in the radius.

We summarise taken parameters in table \ref{tab:3a}.

\begin{table}
\begin{center}
\caption{Model parameters.\label{tab:3a}}
\begin{tabular}{ccccc}
\hline
Model         & $\lambda_0$ & $A_0$ & $n_0$      & $E_0$              \\
              &  (pc)       &       & (cm$^{-3}$)& (erg)              \\
\hline\hline
A\ldots       &  32         & 0.25  & 0.2        & $10^{52}$          \\
B\ldots       &  64         & 0.25  & 0.2        & $10^{52}$          \\
C\ldots       &  32         & 0.125 & 0.2        & $10^{52}$          \\
D\ldots       &  32         & 0.25  & 0.8        & $10^{52}$          \\
E\ldots       &  32         & 0.25  & 0.2        & $2.5\times10^{51}$ \\
\hline
\end{tabular}
\end{center}
\end{table}

\section{Results}\label{sec:result}

\subsection{Dynamical Evolution of SNR}
\label{sec:de2}

First of all, let us give some basic formulae about dynamical
 evolution of SNR. 
The evolution of SNR is divided into the following three phases,
 (1) free expansion phase, (2) Sedov phase, and (3) radiative
 cooling phase. 
The free expansion phase continues until the ejecta sweeps up
 about the same amount of the mass of ejecta $M_{\rm ej}$
 in ISM around SNR. 
The transition radius from free expantion to Sedov phase
 $R_{\rm Sedov}$ and the transition time $t_{\rm Sedov}$ are
 written as
\begin{eqnarray}
R_{\rm Sedov}
\sim 19~{\rm pc}
       \left(\frac{n_0}{0.2~{\rm cm}^{-3}}\right)^{-1/3}
       \left(\frac{M_{\rm ej}}{200~M_{\odot}} \right)^{1/3},
\label{eq:rsedov2}
\end{eqnarray}
\begin{eqnarray}
t_{\rm Sedov}
\sim 8.7 \times 10^3~{\rm yr} 
       \left(\frac{n_0}{0.2~{\rm cm}^{-3}}\right)^{-1/3}
 \left(\frac{M_{\rm ej}}{200~M_{\odot}} \right)^{5/6}
          \left(\frac{E_0}{10^{52}~{\rm erg}}\right)^{-1/2}.
\label{eq:tsedov}
\end{eqnarray}
In the Sedov phase, the expansion of shock wave is well
 approximated by a self-similar solution,
\begin{equation}
R_{\rm bub}
\sim \left(\frac{E_0}{\rho_0}\right)^{1/5} t^{2/5},
\label{eq:sedov_r}
\end{equation}
 where $R_{\rm bub}$ is the shock radius at time $t$
 after the explosion and $\rho_0$ is the average mass
 density of ISM. 
The shock speed is given as the time derivative of
 $R_{\rm bub}$,
\begin{equation}
v_{\rm bub} = \frac{dR_{\rm bub}}{dt}
= \frac{2}{5} \left(\frac{E_0}{\rho_0}\right)^{1/5} t^{-3/5}.
\label{eq:sedov_v}
\end{equation}
When the gas at the shock front becomes radiative,
 a dense shell forms at the outer boundary of SNR. 
The transition into the radiative cooling phase occurs
 at the time $t_{\rm cool}$ and the radius $R_{\rm cool}$
 where
\begin{eqnarray}
t_{\rm cool} \sim 5.1 \times 10^5~{\rm yr} 
\left(\frac{n_0}{0.2~{\rm cm}^{-3}}\right)^{-3/4}
          \left(\frac{E_0}{10^{52}~{\rm erg}}\right)^{1/8},
\label{eq:tcool}
\end{eqnarray}
\begin{eqnarray}
R_{\rm cool} \sim 120~{\rm pc} 
\left(\frac{n_0}{0.2~{\rm cm}^{-3}}\right)^{-1/5}
           \left(\frac{E_0}{10^{52}~{\rm erg}}\right)^{1/5},
\label{eq:rcool}
\end{eqnarray}
 where we assumed the radiative cooling is dominated by
 free-free radiation of H and He \citep{shu80}.
On the other hand, the time scale of the Compton cooling
 for the primordial SNR is given as
\begin{equation}
t_{\rm Comp}
\sim 7 \times 10^6~{\rm yr} 
     \left(\frac{1+z}{20}\right)^{-4}, 
\label{eq:tcomp}
\end{equation}
 which is larger than $t_{\rm cool}$ for $z \lesssim 30$
 in which we are interested. 
As stated above, in this article, we focus on the
 generation of the magnetic fields in the Sedov phase
 ($t_{\rm Sedov} < t < t_{\rm cool}$).

In fact, there is another important time scale at which
 the electrons and ions can be regarded as a one-temperature
 fluid. 
The typical time scale is given by the age at which
 the electron and ion equipartition time just behind the shock
 front $\tau_{\rm eq}$ is sufficiently shorter than the age,
 that is, $\tau_{\rm eq} < 0.1 t$ \citep{cox72,ito78},
\begin{eqnarray}
t_{\rm relax}
\sim 3.3 \times 10^4~{\rm yr} 
     \left(\frac{n_0}{0.2~{\rm cm}^{-3}}\right)^{-4/7}
     \left(\frac{E_0}{10^{52}~{\rm erg}}\right)^{3/14}.
\label{eq:trelax}
\end{eqnarray}
The mean radius of SNR at this time is written as
\begin{eqnarray}
R_{\rm relax}
\sim 45~{\rm pc}
     \left(\frac{n_0}{0.2~{\rm cm}^{-3}}\right)^{-1/5}
     \left(\frac{E_0}{10^{52}~{\rm erg}}\right)^{1/5}.
\label{eq:rrelax}
\end{eqnarray}
At this time, the mean temperature of electrons is 
 $\langle T_e \rangle = 0.92~\langle T \rangle$ where the each
 mean temperature is defined with the ion number density $n$ as
\begin{eqnarray}
\langle T_e\rangle /T_s=\frac{\int_{0}^{R_{\rm bub}}n^2[T_e/T_s]R^2dR}
{\int_{0}^{R_{\rm bub}}n^2R^2dR}, 
\\
\langle T \rangle /T_s=\frac{\int_{0}^{R_{\rm bub}}n^2[T/T_s]R^2 dR}
{\int_{0}^{R_{\rm bub}}n^2R^2dR},
\label{eq:teave}
\end{eqnarray}
 where $T_s$ represents the gas temperature at the shock
 front.
As already explained, the relaxation of the electron
 temperature is crucial for the Biermann mechanism.

The above characteristic time scales and radii for the
 five models adopted here are summarised in tables
 \ref{tab:3b} and \ref{tab:3c}.
Here, $t_{\rm end}$ in table \ref{tab:3b} is the time
 when we stopped the calculation, and $R_{\rm end}$ in
table \ref{tab:3c} is the corresponding radius of SNR.

\begin{table}
\begin{center}
\caption{Time scale for each epoch.\label{tab:3b}}
\begin{tabular}{ccccc}
\hline
Model   & $t_{\rm Sedov}$ & $t_{\rm relax}$ & $t_{\rm end}$ & $t_{\rm cool}$ \\
        & (10$^{3}$ yr)   &  (10$^{4}$ yr)  & (10$^{5}$ yr) & (10$^{5}$ yr)  \\
\hline\hline
A\ldots &  8.7            & 3.3             & 3.4           & 5.0            \\
B\ldots &  8.7            & 3.3             & 3.4           & 5.0            \\
C\ldots &  8.7            & 3.3             & 3.4           & 5.0            \\
D\ldots &  5.5            & 1.5             & 1.7           & 1.8            \\
E\ldots & 17.3            & 2.4             & 4.1           & 4.2            \\
\hline
\end{tabular}
\end{center}
\end{table}

\begin{table}
\begin{center}
\caption{Mean radius for each epoch.\label{tab:3c}}
\begin{tabular}{ccccc}
\hline
Model   & $R_{\rm Sedov}$ & $R_{\rm relax}$ & $R_{\rm end}$ & $R_{\rm cool}$ \\
        & (pc)            & (pc)            & (pc)          & (pc)           \\
\hline\hline
A\ldots & 19              & 45              & 115           &  133           \\
B\ldots & 19              & 45              & 115           &  133           \\
C\ldots & 19              & 45              & 115           &  133           \\
D\ldots & 12              & 25              &  66           &   67           \\
E\ldots & 19              & 30              &  93           &   94           \\
\hline
\end{tabular}
\end{center}
\end{table}

\subsection{Generated Magnetic Fields}

\begin{figure}
\begin{center}
\FigureFile(150mm,0mm){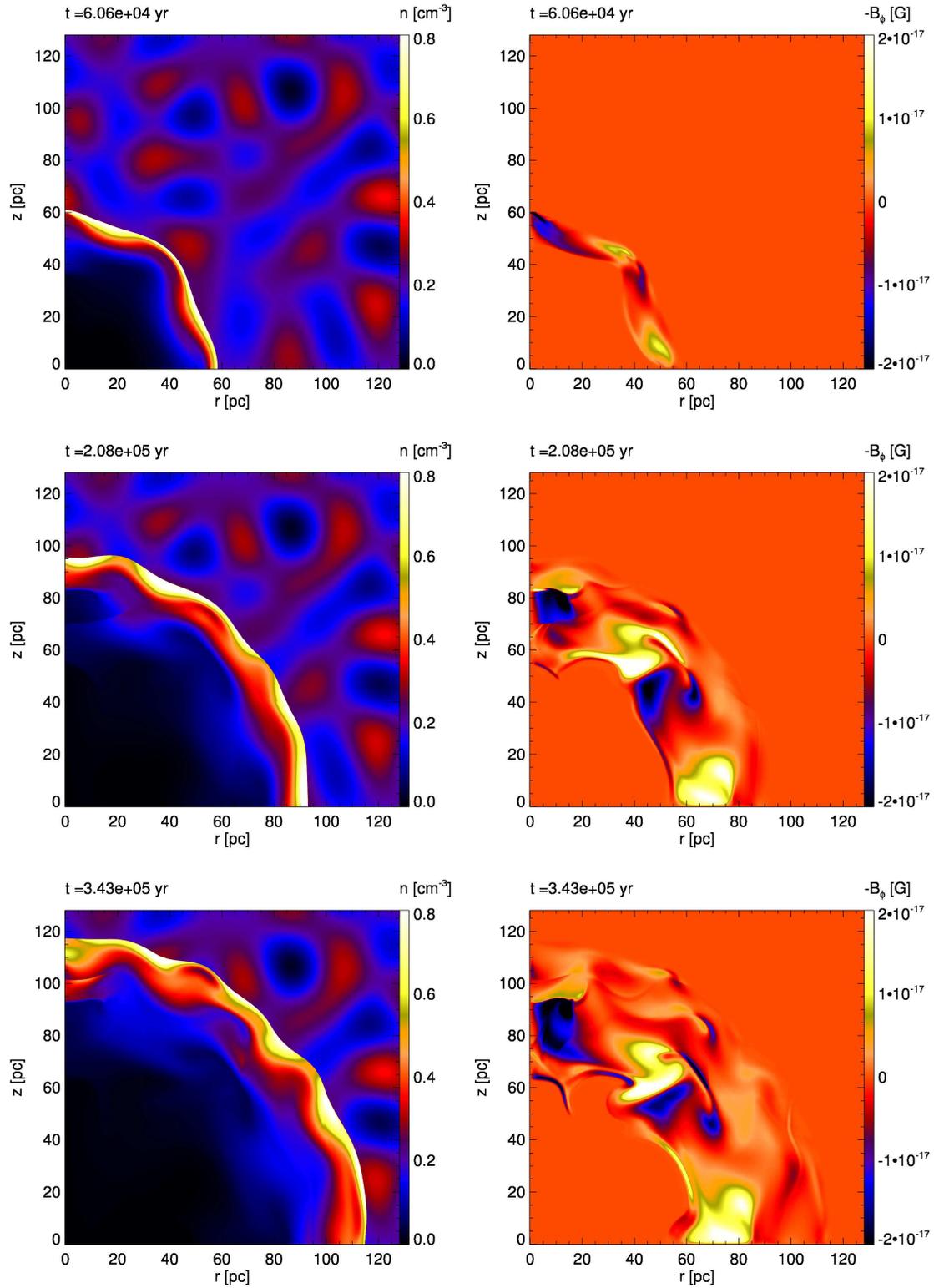}
\end{center}
\caption{Distributions of the gas number density (left) and 
the magnetic fields (right) at 
$t = 6\times10^{4}$ (top panels), $2\times10^{5}$ (middle panels), 
and $3.4\times10^{5}$ (bottom panels) yr for model A.
\label{fig:1}}
\end{figure}

The evolution of the structure of the gas density and magnetic
 fields for model A, our fiducial model, is shown in
 figure \ref{fig:1}. 

When the blast wave expands into the surrounding inhomogeneous
 medium and the electron temperature is sufficiently relaxed, the
 magnetogenesis by the Biermann mechanism starts to work. 
For this model, magnetogenesis starts when the SNR expands to the
 radius of $\sim 45~{\rm pc}$.
At $t = 6 \times 10^4~{\rm yr}$, the radius of the bubble
 reaches $\sim 60~{\rm pc}$ and the anisotropic structure is
 clearly seen, which is induced by the interaction of shock front
 and the density inhomogeneity of the ISM. 
The amplitude of the magnetic field is $\sim 10^{-17}~{\rm G}$
 behind the shock front at this time. 
At $t = 2 \times 10^5~{\rm yr}$, the shock expands to
 $\sim 90~{\rm pc}$, and the magnetic fields are
 $\sim 10^{-18}~{\rm G}$ just behind the shock front while
 they are $10^{-17}~{\rm G}$ for the inner hot cavity.
At $t = 3.4\times10^{5}~{\rm yr}$, the radius reaches
 $\sim 110~{\rm pc}$ and the magnetic fields of
 $\sim 10^{-17}-10^{-18}~{\rm G}$ are distributed from the radius
 of 60 to $110~{\rm pc}$.

\begin{figure}
\begin{center}
\FigureFile(150mm,0mm){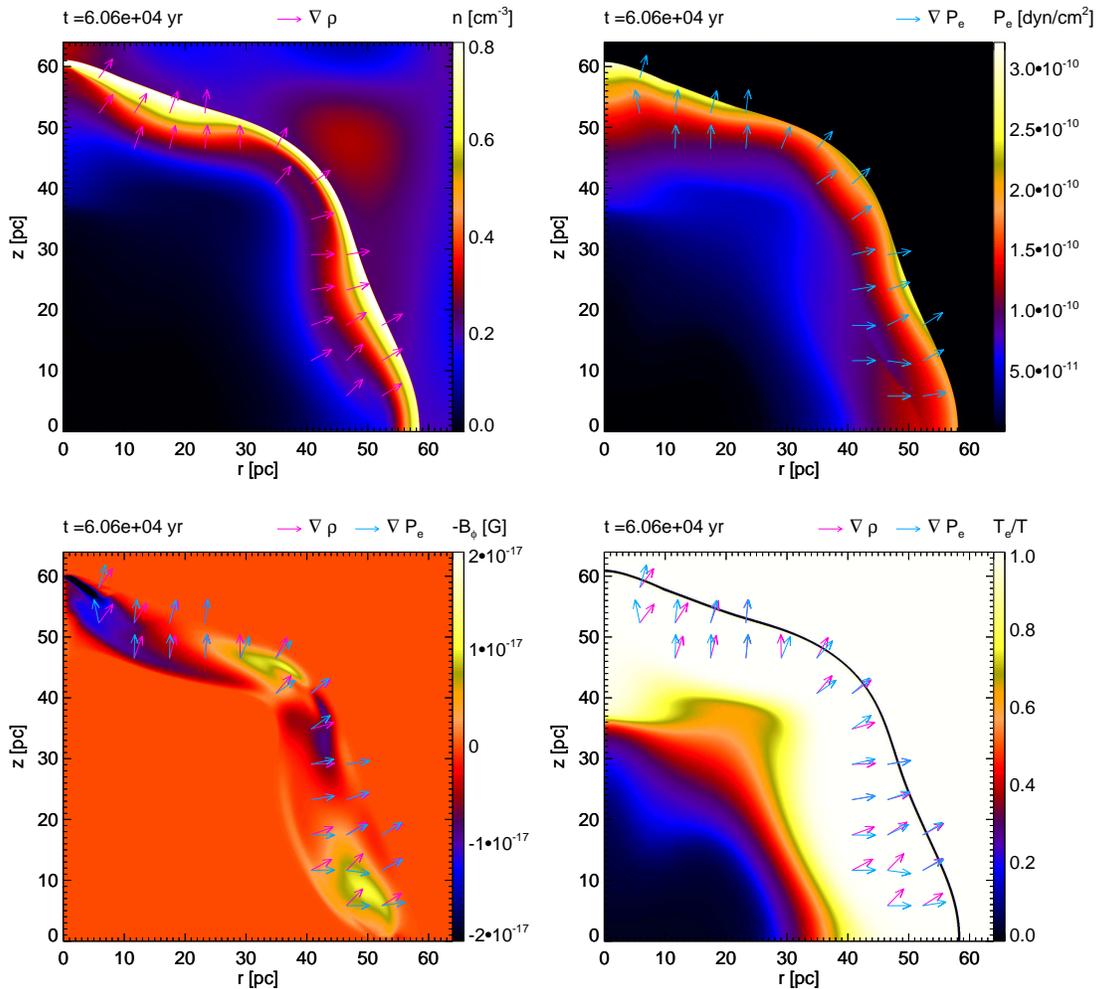}
\end{center}
\caption{Distributions of the gas number density (upper-left), 
 the electron pressure (upper-right), the magnetic fields (lower-left), 
 and the ratio of the electron temperature and the gas temperature 
 (lower-right) at $t = 6 \times 10^{4}~{\rm yr}$ for model A.
Red and blue arrows represent directions of the gradients of 
 the density and the electron pressure, respectively. 
\label{fig:2}}
\end{figure}

Figure \ref{fig:2} shows the distributions of the gas number
 density, the electron pressure, the magnetic fields and the ratio
 of the electron temperature $T_e$ to the gas temperature $T$
 at $t = 6 \times 10^{4}~{\rm yr}$. 
At first glance, the distributions of the density (upper-left panel)
 and the electron pressure (upper-right panel) look very similar. 
However, in fact, the gradient vectors of the density and the
 electron pressure are not exactly parallel (lower panels), which
 is necessary for the Biermann mechanism to work. 
These panels show that the density gradient has the relatively
 larger azimuthal component compared with the pressure gradient. 
This would originate from the density inhomogeneity in ISM. 
This is consistent with the picture of \citet{naka06} where it was
 suggested that the density structure is affected by the
 distribution of the diffuse cloud, while the pressure is mainly
 determined by the global structure of the SNR and has more or less
 radial gradient.
Here it should be important to note that magnetogenesis does not
 occur in the deep interior of the bubble. 
This is because, as we see in the lower-right panel, the electron
 temperature has not been relaxed due to low density.

\begin{figure}
\begin{center}
\FigureFile(140mm,0mm){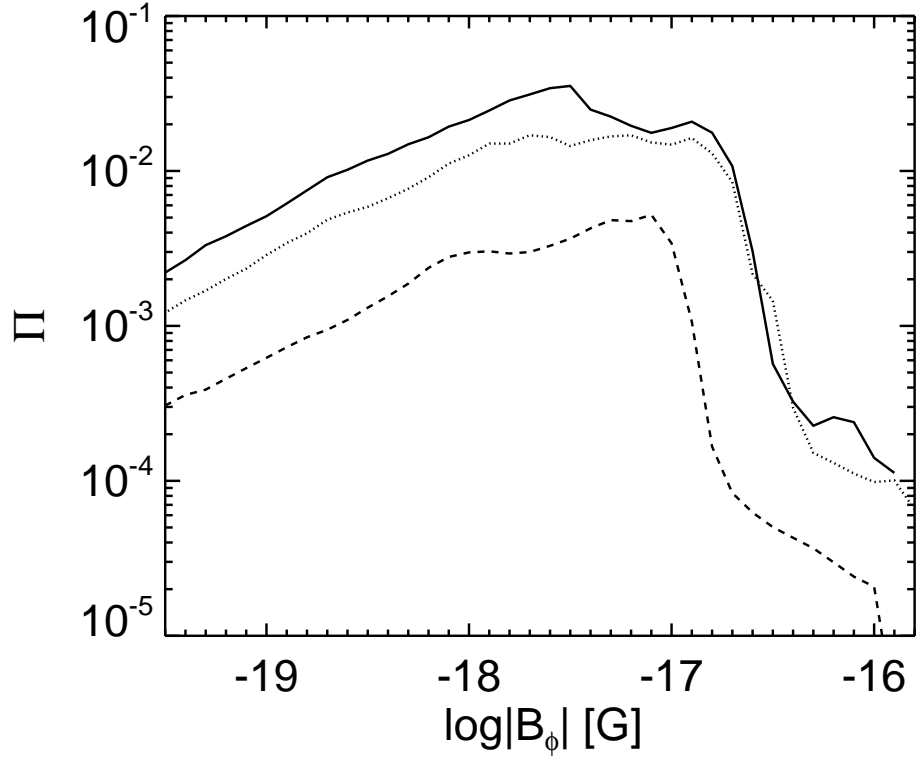}
\end{center}
\caption{
PDF of the strength of the magnetic fields at 
$t = 6\times10^{4}$ (dashed line), $2\times10^{5}$ (dotted line), 
and $3.4\times10^{5}$ (solid line) yr for model A.
\label{fig:3}}
\end{figure}

The probability distribution function (PDF) of the magnetic
 field strength is defined as
\begin{equation}
\Pi(\log |B_{\phi}|) \equiv \frac{N(\log |B_{\phi}|)}{N_{\rm cells}},
\end{equation}
 where $N(\log |B_{\phi}|)$ is the number of cells with magnetic
 fields $B_{\phi}$ and $N_{\rm cells} = 2048^2$ is the total number
 of cells. 
The PDFs at these epochs are shown in figure \ref{fig:3}. 
We can see that the magnetic fields of
 $\sim 10^{-17}-10^{-18}~{\rm G}$ are generated at each epoch.
The magnetic fields of $\sim 10^{-16}~{\rm G}$ are generated
 in restricted number of cells, although the population is not so
 large.

\begin{figure}
\begin{center}
\FigureFile(140mm,0mm){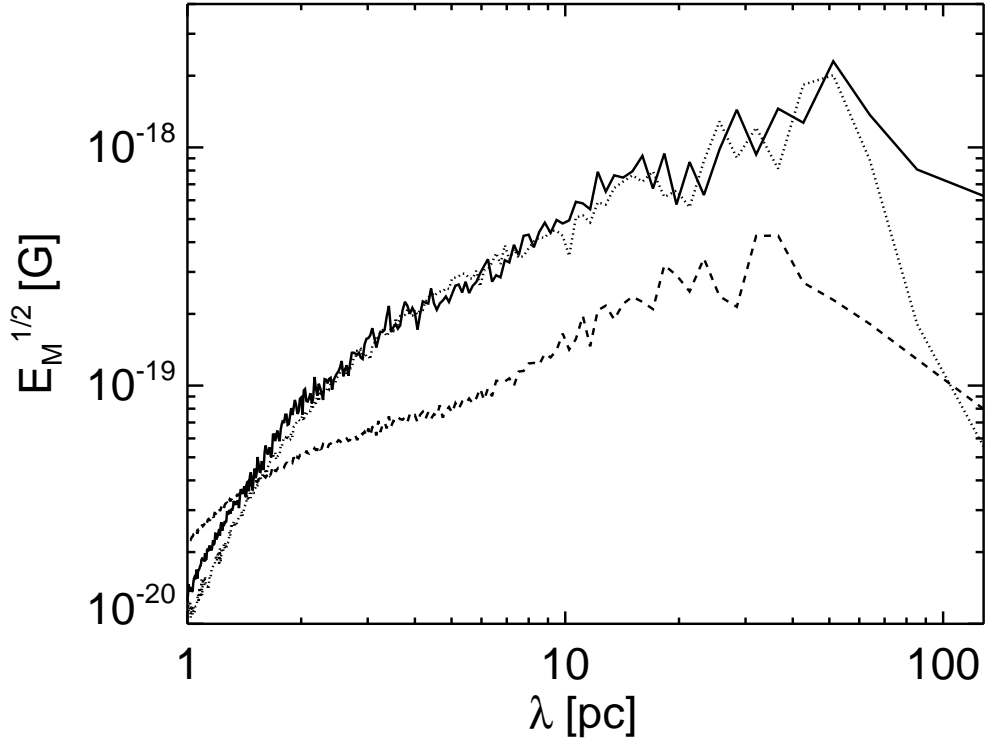}
\end{center}
\caption{
Square root of the energy spectrum of the magnetic fields at 
$t = 6\times10^{4}$ (dashed line), $2\times10^{5}$ (dotted line), 
and $3.4\times10^{5}$ (solid line) yr for model A.
\label{fig:4}}
\end{figure}

Figure \ref{fig:4} shows the square root of the energy spectrum
 of the magnetic fields for several times. 
The energy spectrum of the magnetic fields is calculated by using
 a definition of shell-averaged magnetic power spectrum $P_M(k)$
 derived by \cite{chr01}. 
Fourier series ${\ve B_f}(k_{r},k_{z})$ is written as
\begin{equation}
{\ve B_f}(k_{r},k_{z})
= \int_{-L}^{L}\int_{-L}^{L}
  \frac{drdz}{(2L)^2}
  {\ve B}(r,z) 
  e^{-i \frac{\pi}{L} (k_r r + k_z z)},
\end{equation}
 where $k_{r}$, and $k_{z}$ represent wave numbers for respective 
 directions in a period $2L~(=256~{\rm pc})$. 
We assume ${\ve B} = (0, B_{\phi}, 0)$ in the first quadrant, and
 the antisymmetric one against the axis in other quadrants. 
The power spectrum $P_M(k)$ in Fourier space is defined as 
\begin{eqnarray}
P_M(k)=\langle \ve{B}^*_f \cdot \ve{B}_f \rangle,
\end{eqnarray}
 where $\langle \ve{B}^*_f \cdot \ve{B}_f \rangle$ is the value
 averaged over the shells with constant $k = |{\ve k}|$.
Then, the shell-integrated magnetic energy spectrum $E_M(k)$ is
 written as
\begin{equation} 
E_M(k) = 2\pi k P_M(k).
\label{fig:pm3}
\end{equation}
In figure \ref{fig:4}, $\sqrt{E_M(\lambda)}$ is plotted against
 the wave length $\lambda = 2L/k$. 
At $t = 6 \times 10^{4}~{\rm yr}$, the peak value is
 $\sim 4\times 10^{-19}~{\rm G}$, and the corresponding wave length
 $\lambda_{\rm BP}$ is $37~{\rm pc}$ which is near the scale of the
 initial fluctuation of ISM, $\lambda_0 = 32~{\rm pc}$. 
At $t = 2 \times 10^{5}~{\rm yr}$ and $3.4 \times 10^{5}~{\rm yr}$,
 the spectra are almost identical.
The energy spectrum takes the maximum
 at $\lambda_{\rm BP} = 51~{\rm pc}$ which is nearly equal to
 $1.5 \lambda_0$. 
We can also see that the generation of magnetic fields completed by
 $t = 2 \times 10^5~{\rm yr}$.

The size of the magnetic loop extending in $\phi$-direction equals
 to the typical coherence length appeared in figure \ref{fig:4},
 $\sim \lambda_{\rm BP}/2$. 
Then, the radius of the toroidal fields $r_t$ is considered to be
 $r_t = \lambda_{\rm BP} / 4 \sim 1.5 \lambda_0 / 4$, and the
 coherence length is estimated as
 $2 \pi r_t \sim 1.5 \lambda_{\rm BP} \sim 2\lambda_0
 \sim 64~{\rm pc}$ at $3.4 \times 10^{5}~{\rm yr}$.
This indicates that the magnetic field is expected to be observed
 with a 3-dimensional size of $2 \pi r_t \sim 64~{\rm pc}$ from
 our 2-dimensional simulations.

\subsection{Comparison of Models}

\begin{figure}
\begin{center}
\FigureFile(140mm,0mm){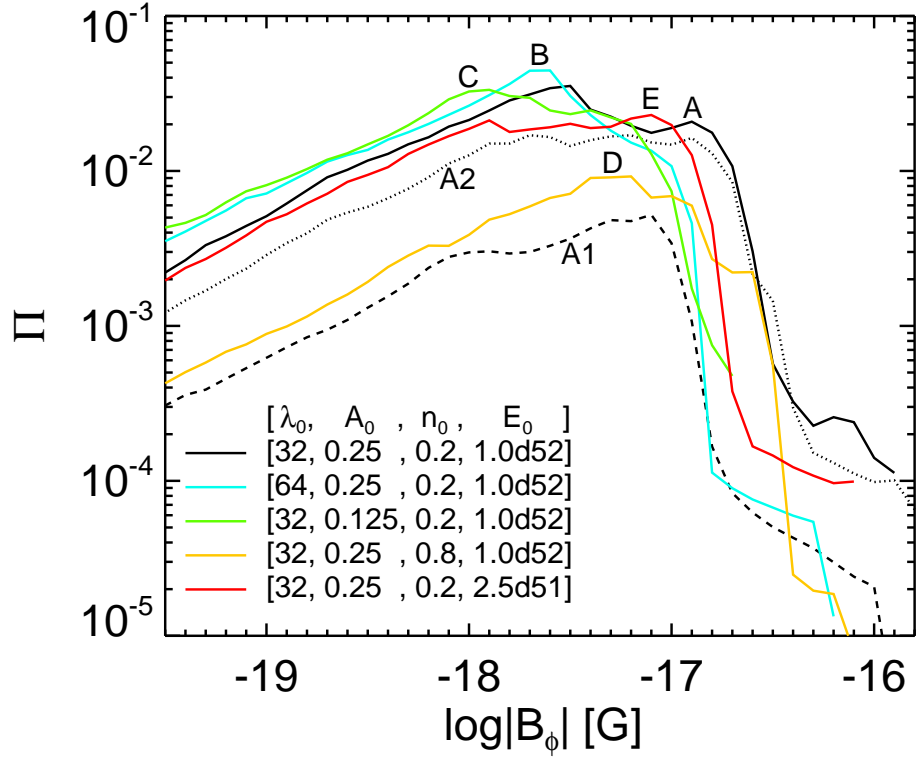}
\end{center}
\caption{
Comparison of PDFs of the strength of the magnetic fields for each model.
Black, blue, green, yellow, 
and red lines represent the models A ($t = 3.4\times10^{5}~{\rm yr}$), 
B ($t = 3.4\times10^{5}~{\rm yr}$), 
C ($t = 3.4\times10^{5}~{\rm yr}$), 
D ($t = 1.2\times10^{5}~{\rm yr}$), 
and E ($t = 4.1 \times 10^{5}~{\rm yr}$), respectively.
Dashed (A1) and dotted (A2) lines represent $t = 6\times10^{4}$ and 
$2\times10^{5}~{\rm yr}$ for model A. 
\label{fig:5}}
\end{figure}

In this subsection, we make a comparison of the results of the five
 models (A-E) shown in table \ref{tab:3a}.
The PDFs of the magnetic field
 strength for the five models are shown in figure \ref{fig:5}.
This indicates that the amplitude of magnetic fields are not
 so different for the variation of our models. 
Considering the maximum amplitude of $|B_{\phi}|$ with
 $\Pi > 10^{-2}$, $|B_{\phi}(\Pi > 10^{-2})|$ of model A is
 roughly twice as large as that of models B and C. 
The radius and large-scale structure for models
 D ($R = 60~{\rm pc}$, $t = 1.2 \times 10^5~{\rm yr}$) and E
 ($R = 90~{\rm pc}$, $t = 4.1 \times 10^5~{\rm yr}$) are similar to
 those of model A at $t = 6 \times 10^4~{\rm yr}$ (A1, dashed line;
 $R = 60~{\rm pc}$) and $2 \times 10^5~{\rm yr}$ (A2, dotted line;
 $R = 90~{\rm pc}$), respectively.
This simply comes from the self-similar evolution 
 of the SNR in the adiabatic phase.
Accordingly, we compare the PDF plots of models D and E 
 with curves A1 and A2, respectively.
Then, for model D, the PDF of the magnetic fields can be compared
 with that of model A at $t = 6 \times 10^4~{\rm yr}$
 (A1, dashed line). 
$|B_{\phi}(\Pi > 10^{-3})|$ of model D at
 $t = 1.2 \times 10^5~{\rm yr}$ is larger than that of model A
 at $t = 6 \times 10^4~{\rm yr}$.
This difference comes from the fact that the post shock pressure
 of model D at $t_{\rm relax} = 1.5 \times 10^4~{\rm yr}$ is larger
 than that of model A at $t_{\rm relax} = 3.3 \times 10^4~{\rm yr}$
 because the amplitude of the generated magnetic fields is
 fundamentally proportional to the pressure gradient. 
The same argument can be applied to the comparison of model E and
 model A at $t = 2 \times 10^5~{\rm yr}$ (A2, dotted line). 
For the population of $\Pi > 10^{-2}$, the amplitude of
 model E is less than that of model A at
 $t = 2 \times 10^5~{\rm yr}$. 
This difference also comes from the fact that the post shock
 pressure of model E at $t_{\rm relax} = 2.4 \times 10^4~{\rm yr}$
 is less than that of model A at
 $t_{\rm relax} = 3.3 \times 10^4~{\rm yr}$.

\begin{figure}
\begin{center}
\FigureFile(140mm,0mm){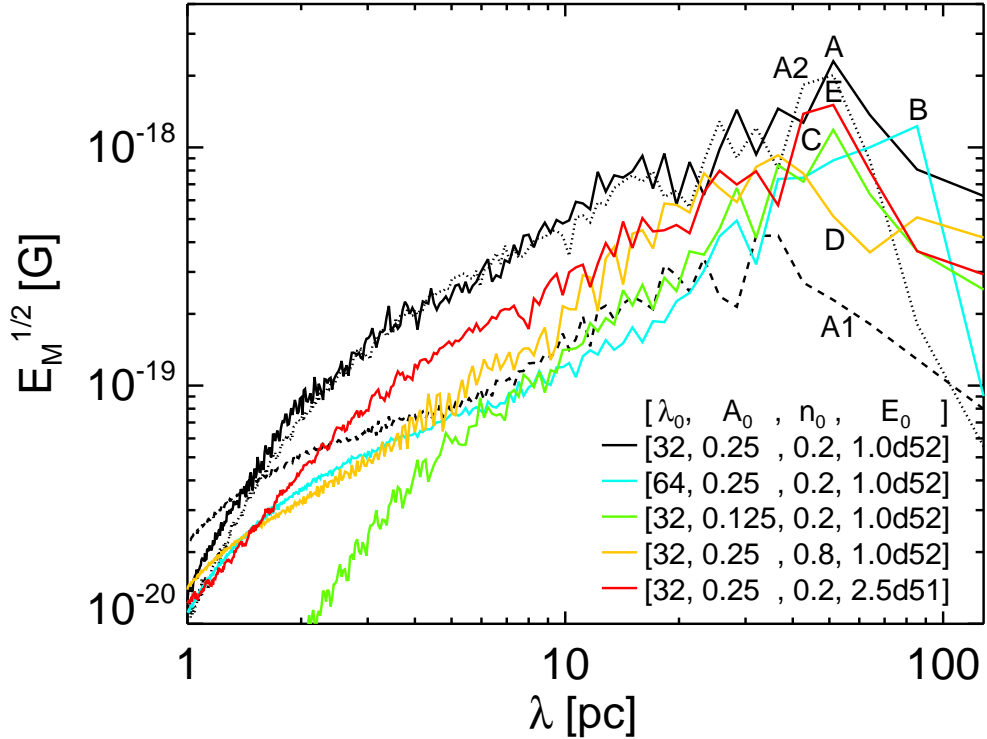}
\end{center}
\caption{
Comparison of the square roots of the energy spectrum of the magnetic 
fields for each model. 
Black, blue, green, yellow, 
and red lines represent models A ($t = 3.4\times10^{5}~{\rm yr}$), 
B ($t = 3.4\times10^{5}~{\rm yr}$), 
C ($t = 3.4\times10^{5}~{\rm yr}$), 
D ($t = 1.2\times10^{5}~{\rm yr}$), 
and E ($t = 4.1\times10^{5}~{\rm yr}$), respectively.
Dashed (A1) and dotted (A2) lines represent $t = 6 \times 10^{4}$ and 
$2 \times 10^{5}~{\rm yr}$ for model A. 
\label{fig:6}}
\end{figure}

Figure \ref{fig:6} shows the comparison of the square roots of
 the energy spectrum of the magnetic fields for the five models.
For models A, C, and E, the curves take their maxima at
 $\sim 51~{\rm pc}$. 
For models B and D, the peak sizes are 85 and $37~{\rm pc}$,
 respectively. 
This shows each spectrum has a peak near the scale of
 $1.5 \lambda_0$ except for high-density model D. 
The peak magnitude of model A is $10^{-18}~{\rm G}$,
 and that is twice as large as that of models B and C.
For high-density model D, we can compare the result of model A at
 $t = 6 \times 10^4~{\rm yr}$ (A1, dashed line), both of which
 have a similar peak size. 
The peak wavelength of both models is $37~{\rm pc}$, although
 the amplitude of the magnetic fields of model D is twice as large
 as that of model A (A1). 
This difference seems to come from the fact that the magnetic
 fields are generated even in the inner region for model D
 since $t/t_{\rm relax}$ is larger compared with other models. 
This is also seen in the comparison of model E at
 $t = 4.1 \times 10^5~{\rm yr}$ and model A at
 $t = 2 \times 10^5~{\rm yr}$ (A2, dotted line). 
The place of the peak wavelength of both models is $51~{\rm pc}$,
 and the peak energy density of model E is slightly lower than
 that of model A (A2).

For the coherence length, if the scale of the maximum energy
 density $\lambda_{\rm BP}$ is equal to $1.5\lambda_0$,
 the 3-dimensional length of toroidal magnetic field is estimated
 as $2\pi \lambda_{\rm BP}/4 \sim 2\lambda_0$. 
This means that the coherence length of the toroidal field of
 $10^{-18}~{\rm G}$ is estimated as $\sim 64~{\rm pc}$ for
 models A, C, and E, while it is $\sim 128~{\rm pc}$ and
 $\sim 32~{\rm pc}$ for models B and D, respectively. 

\begin{figure}
\begin{center}
\FigureFile(140mm,0mm){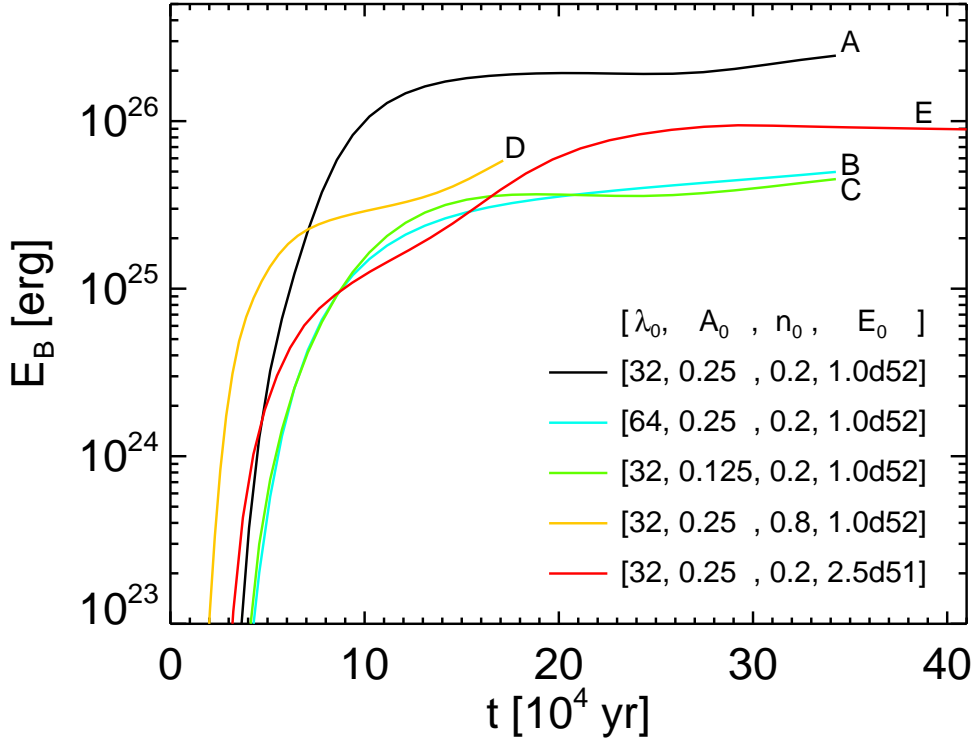}
\end{center} 
\caption{Time evolution of the total magnetic energy produced by the
 Biermann mechanism for various models.
Black, blue, green, yellow,  
and red lines represent models A, B, C, D, and E, respectively.
\label{fig:7}}
\end{figure}

Time evolution of the total magnetic energy is shown in
 figure \ref{fig:7}. 
The final magnetic energy for each model indicated by the figure is,
 $\sim 2 \times 10^{26}~{\rm erg}$ for model A, 
 $\sim 5 \times 10^{25}~{\rm erg}$ for model B, 
 $\sim 4 \times 10^{25}~{\rm erg}$ for model C, 
 $\sim 6 \times 10^{25}~{\rm erg}$ for model D and
 $\sim 9 \times 10^{25}~{\rm erg}$ for model E, respectively.
Apparent knees around $5 \times 10^4 - 2 \times 10^5~{\rm yr}$
 come from the time scale of the electron temperature equilibrium
 shown in table \ref{tab:3b}. 
The total magnetic energy of model A is several times larger than
 those of other models.
This behavior will be interpreted by an analytical estimation of
 the magnitude of the magnetic fields and the time evolution of
 the total energy in the next section. 
Although there are many uncertainties in our initial conditions,
 it is implied that the amplitude of the generated magnetic fields
 does not depend on SNR and ISM parameters so strongly.

\begin{figure}
\begin{center}
\FigureFile(140mm,0mm){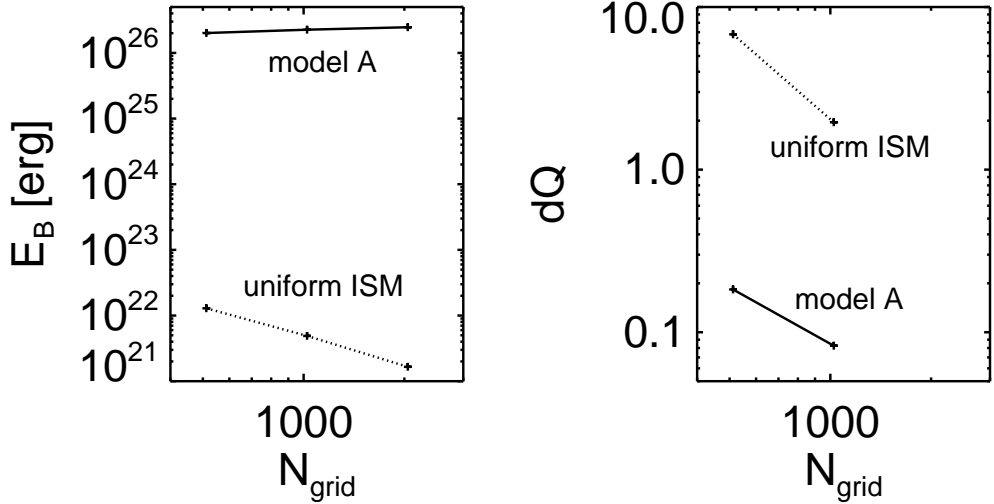}
\end{center}
\caption{
Convergence study of total magnetic energy $E_B$ (left panel) at
 $3.4 \times 10^5~{\rm yr}$.
$dQ$ (right panel) means the relative convergence error in $E_B$ 
of each model compared with the highest resolution model 
$N_{\rm grid} = 2048$, defined as 
$dQ=|E_B(N_{\rm grid})-E_B(N_{\rm grid}=2048)|/E_B(N_{\rm cl}=2048)$.
$N_{\rm grid}$ represents the number of grid points in one dimension 
and the spatial resolution of the simulation. 
Solid line means the result of the models which contain 
the inhomogeneity (model A), while dashed line means that of
 the models without the density fluctuation (uniform ISM).
\label{fig:8}}
\end{figure}

Finally, let us argue the convergence of the results of our
 simulations. 
Figure \ref{fig:8} shows a convergence study in comparison
 with the simulations of lower spatial resolution
 in which the numerical box size is taken identically but
 the grid points are reduced as $N_{\rm grid} = 512$ and 1024.
$E_{B}$ and $dQ$ represent the total magnetic energy of SNR at
 $3.4 \times 10^5~{\rm yr}$ and the relative error in $E_B$ of
 each low-resolution model compared with that of the highest
 resolution model $N_{\rm grid} = 2048$, defined as 
 $dQ=|E_B(N_{\rm grid}) - E_B(N_{\rm grid}=2048)|
 / E_B(N_{\rm grid}=2048)$. 
In both panels, solid line is the result of the calculations of
 the interaction between the inhomogeneous ISM and SNR (model A)
 and dashed line is that of the evolution of SNR in uniform ISM.
Left panel could be a measure of the numerical error in magnetic
 fields of our simulations. 
We can see that, in the highest resolution model, $E_B$ of the
 model with homogeneous ISM is about 10$^5$ times smaller than 
 that of model A and the numerical error of the magnetic fields
 generated by the curvature effect in the interior of SNR is
 negligible. 
In right panel, solid line shows that our simulation is almost
 converged and the difference is about 10\% for the lower
 resolution model of $N_{\rm grid} = 1024$. 
This indicates that roughly 128 grids for the radius of density
 fluctuation $r_0 = \lambda_0/4 = 8~{\rm pc}$ are required for a
 converged calculation.

\section{Discussion}\label{sec:discuss}

\subsection{Characteristics of Magnetic Fields}

In this subsection we give an order-of-magnitude estimation of
 the amplitude of the magnetic fields and the time evolution of
 the total magnetic energy generated by the Biermann mechanism.
We extend the analysis in \cite{han05} considering the time
 evolution of the physical quantities of the SNR bubble and
 the relaxation of the electron temperature. 
The characteristic pressure at the shock front is given by the ram
 pressure, $P \sim P_{\rm ram} \sim (3/4) \rho_0 v_{\rm bub}^2$,
 and its gradient is estimated as $\nabla P \sim P/L$
 where the pressure scale height can be evaluated as a typical
 shell width $L \sim R_{\rm bub}/10$. 
Noting that the density gradient is determined by the fluctuation
 of ISM, we have
 $\nabla \rho \sim A_0 \rho_0/(2 r_0) = 2 A_0 \rho_0/\lambda_0$.
Thus, the amplitude of magnetic fields generated within
 a characteristic time, $\tau \sim L/v_{\rm bub}$,
 is given by (see equation (\ref{ind2}))
\begin{eqnarray}
B_{\rm Bier}
&=& \left| \alpha
           \frac{\nabla P \times \nabla \rho}{\rho^2} \tau
    \right|
\nonumber \\
&\sim& \alpha \frac{3 v_{\rm bub} A}{2 \lambda_0}
\sim \frac{3 \alpha A}{5 \lambda_0}
     \left(\frac{E_0}{\rho_0}\right)^{1/5} t^{-3/5}
\nonumber \\
&\sim& 10^{-17}~{\rm G} 
       \left(\frac{\lambda_0}{32~{\rm pc}}\right)^{-1}
       \left(\frac{A}{0.5}\right)
       \left(\frac{n_0}{0.2~{\rm cm}^{-3}}\right)^{-1/5}
\nonumber \\
& &    \times
       \left(\frac{E_0}{10^{52}~{\rm erg}}\right)^{1/5}
\left(\frac{t}{10^5~{\rm yr}}\right)^{-3/5},
\label{eq:B-estimate}
\end{eqnarray}
 where we used equation (\ref{eq:sedov_v}) and put
 $\alpha \sim 0.5 \times 10^{-4}~{\rm G~sec}$. 
This is reasonably consistent with the value obtained from our
 numerical simulations.

\begin{figure}
\begin{center}
\FigureFile(140mm,0mm){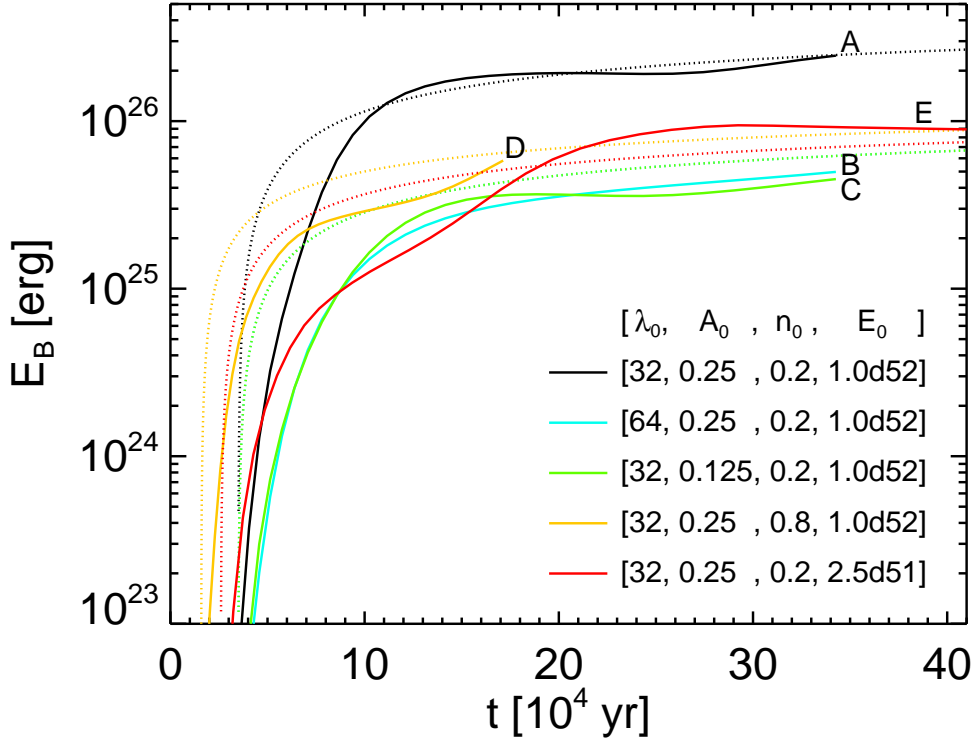}
\end{center} 
\caption{The same as figure \ref{fig:7} but added lines 
of the analytic estimation obtained from equation (\ref{eq:eb3}). 
Black, blue, yellow, orange and red lines represent 
models A, B, C, D, and E, respectively.
Solid and dotted lines represent the numerical result and the analytic 
estimation.
\label{fig:9}}
\end{figure}

On the other hand, the growth of the total magnetic energy during
 a time interval in which an SNR expands from volume $V$ to $V+dV$
 can be estimated as
\begin{eqnarray}
dE_B 
&\sim& \frac{B_{\rm Bier}^{2}}{8 \pi} dV
\sim \frac{B_{\rm Bier}^{2}}{8 \pi} 4\pi R^2 dR 
\nonumber \\
&\sim& \frac{9\alpha^2 A^2 E_0}{125\lambda_0^2 \rho_0}t^{-1}dt,
\label{eq:E-estimate2}
\end{eqnarray}
 where $dV (= 4 \pi R^2 dR)$ is a difference in the volume of SNR
 between two epochs $t$ and $t+dt$ and we used equation
 (\ref{eq:sedov_r}).
The total magnetic energy $E_B(t)$ contained in a SNR of the age $t$
 is given by the time integration of $dE_B(t)/dt$ from
 $t_{\rm start}$ to $t$. 
Assuming $t_{\rm start} \sim t_{\rm relax}$ for the sufficient
 equilibrium of the inner region, we obtain the total magnetic
 energy as 
\begin{eqnarray}
E_B(t)
&\sim& \int_{ t_{\rm start} }^{ t } 
       \frac{9\alpha^2 A^2 E_0}{125\lambda_0^2 \rho_0}t^{-1}dt
\nonumber \\
&=& \frac{9\alpha^2 A^2 E_0}{125\lambda_0^2 \rho_0}
    (\ln{t} - \ln{t_{\rm start}})
\nonumber \\
&\sim& 3 \times 10^{26}~{\rm erg} 
       \left(\frac{\lambda_0}{32~{\rm pc}}\right)^{-2}
       \left(\frac{A}{0.5}\right)^2
       \left(\frac{n_0}{0.2~{\rm cm}^{-3}}\right)^{-1}
\nonumber \\
&& \times
       \left(\frac{E_0}{10^{52}~{\rm erg}}\right)
      \ln{\left(\frac{t/t_{\rm relax}}{10}\right)}.
\label{eq:eb3}
\end{eqnarray}
See table \ref{tab:3b}. 
This analytic estimation is also plotted in figure
 \ref{fig:9} to compare with our numerical simulation. 
Consistency between these estimations and the numerical results is
 remarkable, explaining not only the qualitative behavior but also
 the absolute magnitudes.

\subsection{Implication for Seed Magnetic Fields}
\label{sec:ismf3}

Now we estimate the spatially-averaged energy density of the
 magnetic fields in protogalaxies expected from the first
 star SNR and consider whether they could be a source of the seed
 fields or not.

The total number density of the SNe is roughly estimated as
\begin{equation}
n_{\rm SN}
\simeq \frac{\dot{\rho}_{\star, {\rm III}}\tau}{M_{\rm S}},
\label{eq:nsn}
\end{equation}
 where $\dot{\rho}_{\star, {\rm III}}$, $\tau$, and $M_{\rm S}$
 are the primordial star formation rate (SFR) of Pop III stars
 per unit volume, the duration of the first star formation and
 a typical mass of the first stars, respectively. 
As for the SFR of Pop III stars, extrapolating the one by
 \citet{pel04} and \citet{ric04}, we have
\begin{equation}
\dot{\rho}_{\star, {\rm III}}
\sim 6 \times 10^{-4}~M_{\odot}~{\rm yr}^{-1}~{\rm Mpc}^{-3}
     \left( \frac{f_{\rm III}}{0.06} \right),
\end{equation}
 where $f_{\rm III}$ is the fraction of the Pop III stars in SFR,
 and we adopt $f_{\rm III} = 0.06$ that was derived under the
 assumption that very massive black holes produced from first stars
 end up in supermassive black holes in galactic centers
 \citep{sch02}. 
If we assume the formation period of the first stars continued from
 $z\sim 20$ to $10$ ($\tau \sim 0.3~{\rm Gyr}$), the total number
 density of the Pop III SNe can be estimated as
\begin{eqnarray}
n_{\rm SN}  
&\sim& 4 \times 10^{-68}~{\rm cm^{-3}}
       \left(\frac{1+z}{11}\right)^3
       \left(
       \frac{\dot{\rho}_{\star, {\rm III}}}{6 \times 10^{-4}~
             M_{\odot}~{\rm yr}^{-1}~{\rm Mpc}^{-3}}
       \right)
\nonumber \\
&&     \times
       \left(\frac{\tau}{0.3~{\rm Gyr}}\right)
       \left(\frac{M_{\rm S}}{200~M_{\odot}} \right)^{-1},
\label{eq:nsn3a}
\end{eqnarray}
 where the number density is in units of physical scale,
 not comoving. 
Primordial star formation rate was also estimated by \citet{gre06}. 
Even taking Pop III and Pop II.5 of their classification into
 account, we confirmed that our results below do not change so much.

Taking the typical value of the magnetic energy of a Pop III SNR
 as that of model A, $E_{B} \sim 10^{26}~{\rm erg}$,
 the spatially-averaged magnetic energy density is estimated as,
 $e_B \sim 10^{-42}~{\rm erg~cm^{-3}}$. 
If we assume that galaxies are formed in such a magnetized medium,
 the magnetic energy density in protogalaxies is given by
\begin{eqnarray}
e_{B, {\rm gal}}
&\sim& e_B\Delta^{4/3} \nonumber \\
&\sim& 10^{-39}~{\rm erg~cm^{-3}}
       \left(\frac{\Delta}{200}\right)^{4/3}
       \left(\frac{e_B}{10^{-42}~{\rm erg~cm^{-3}}}\right), 
\label{eq:11}
\end{eqnarray}
 where $\Delta$ represents the overdensity of protogalaxies.
This means that the average magnitude of the magnetic fields becomes 
 $B \sim 10^{-19}~{\rm G}$, which would be enough for the required
 seed field of galactic dynamo \citep{les95}. 
Although the coherence length of the order of $10-100~{\rm pc}$
 estimated here is much smaller than the galactic scale, it would
 be amplified by the galactic dynamo \citep{poe93,bec94,bec96}. 
It may also be amplified by interstellar turbulence dynamo to
 produce fluctuating components \citep{bal04}. 

Finally let us comment on the three-dimensional effects.
In this study, we performed two-dimensional MHD simulations.
However, because of the assumption of the axisymmetry,
 the generated magnetic fields are restricted to the toroidal
 component. 
This makes it rather hard to argue the coherence length of magnetic
 fields. 
Further, the spectrum of magnetic fields would be different in
 three-dimensional simulations, because the vorticity cascade is
 different in 2D and 3D turbulences.
We will present three-dimensional simulations in a separate paper
 but we believe that most of the features of the Biermann mechanism
 in SNR are captured in the present study.

\section{Summary}

In this article, we argued the Biermann mechanism in primordial
 supernova remnants through two-dimensional MHD simulations with
 the Biermann term. 
We solved simultaneously the relaxation of the electron
 temperature, which is crucial to the efficiency of the Biermann
 mechanism and was not taken into account in our previous study
 \citep{han05}. 
It was found that magnetic fields begin to be generated from
 $t = t_{\rm relax}$ just behind the shock front. 
The total magnetic energy reaches about $10^{26}~{\rm erg}$ and
 does not depend strongly on the parameters of SNR and ISM. 
We could understood analytically the dependence of the magnetic
 total energy on the parameters and also the time evolution. 
Finally we evaluated the expected amplitude of magnetic fields in
 protogalaxies, which would be sufficient for seed fields of the
 observed galactic magnetic fields.

\bigskip

HH would like to express sincere thanks to Prof. Tomoyuki Hanawa,
 Dr. Tomoaki Matsumoto, Dr. Kazuya Saigo, Dr. Dai G. Yamazaki,
 and Dr. Motohiko Kusakabe for helpful advice on technical problems
 in numerical works and encouragement.
HH also thanks Prof. Ryoji Matsumoto and Dr. Takaaki Yokoyama for
 a contribution to the calculation code, CANS (Coordinated 
 Astronomical Numerical Software). Numerical computations were
 carried out on Cray XT4 and NEC SX9 systems at the Center for
 Computational Astrophysics of NAOJ, and NEC SX8 system at Yukawa
 Institute for Theoretical Physics, Kyoto University.
Keitaro Takahashi is supported in part by MEXT Grant-in-Aid for
 the global COE programs "Quest for Fundamental Principles in
 the Universe: from Particles to the Solar System and the Cosmos"
 at Nagoya University.
A part of this work (Tomisaka) was supported from Grant-in-Aid
 for Scientific Research (17340059) from MEXT.


\clearpage


\end{document}